\if0\documentclass[reprint,preprintnumbers,
amsmath,amssymb,aps,nofootinbib,superscriptaddress]{revtex4-1}
\usepackage{dcolumn}
\usepackage{bm}
\usepackage{natbib}
\usepackage{cases}
\usepackage{mathrsfs}
\usepackage{hyperref}
\usepackage{color}\fi

\documentclass[reprint,preprintnumbers,
amsmath,amssymb,aps,nofootinbib,superscriptaddress]{revtex4-1}
\usepackage[pdftex]{graphicx}
\usepackage{dcolumn}
\usepackage{bm}
\usepackage[pdftex]{hyperref}
\usepackage[svgnames]{xcolor}
\definecolor{phthaloblue}{rgb}{0.0, 0.06, 0.54}
\definecolor{reddish}{rgb}{0.65, 0.2, 0.2}
 \hypersetup{
    colorlinks=true,
    linkcolor=reddish,
    citecolor=reddish,
    filecolor=reddish,
    urlcolor=reddish,
    }
\makeatletter \def\@eqnnum{{\normalsize \normalcolor (\theequation)}}  \makeatother %

\newcommand{\beeq}{\begin{equation}}
\newcommand{\eneq}{\end{equation}}
\newcommand{\beal}{\begin{align}}
\newcommand{\eeal}{\end{align}}
\newcommand{\ef}{\end{fleqn}}

\newcommand{\nn}{\nonumber\\}

\newcommand{\vep}{\varepsilon}

\newcommand{\Lmd}{\Lambda}

\newcommand{\be}{\begin{equation}}
\newcommand{\ee}{\end{equation}}
\newcommand{\bea}{\begin{eqnarray}}
\newcommand{\eea}{\end{eqnarray}}

\newcommand{\brkt}[1]{\left( #1 \right)}
\newcommand{\brc}[1]{\left\{ #1 \right\}}
\newcommand{\sbk}[1]{\left[ #1 \right]}

\newcommand{\cO}{{\cal O}}

\begin{document}
\preprint{\begin{minipage}[b]{1\linewidth}
\begin{flushright}KEK-TH-2040\\ 
TU-1066\end{flushright}
\end{minipage}}
\title{Revisiting regularization with Kaluza-Klein states and Casimir vacuum 
energy from extra dimensional spaces}
\author{Hiroki Matsui}
\email{hiroki.matsui.c6@tohoku.ac.jp}
\affiliation{Department of Physics, Tohoku University, Sendai, 980-8578 Japan}
\author{Yoshio Matsumoto}
\email{yoshio@cc.miyakonojo-nct.ac.jp}
\affiliation{National Institute of Technology, Miyakonojo College, Miyakonojo, Miyazaki 885-0006, Japan}

\begin{abstract}
In the present paper, we investigate regularization of the one-loop quantum corrections
with infinite Kaluza-Klein (KK) states and evaluate Casimir vacuum energy from extra dimensions.
The extra dimensional models always involve the infinite massless or massive Kaluza-Klein states,
and therefore, the regularization of the infinite KK corrections is highly problematic.
In order to avoid the ambiguity, we adopt the proper time integrals and 
the Riemann zeta function regularization in evaluating the summations of infinite KK states.
In the calculation, we utilize the KK regularization method with exchanging the infinite summations and the infinite loop integrals. At the same time, we also evaluate the correction by the the dimensional regularization method without exchanging the summations and the loop integrals.
Then, we clearly show that the regularized Casimir corrections from the KK states 
have the form of $\propto 1/R^2$ for the Higgs mass and $\propto 1/R^4$ 
for the cosmological constant, where $R$ is the compactification radius.
We also evaluate the Casimir energy in supersymmetric extra-dimensional models. 
The contributions from bulk fermions and bulk bosons 
are not offset because we choose SUSY breaking boundary conditions.
The non-zero supersymmetric Casimir corrections from extra dimensions 
undoubtedly contribute to the Higgs mass and the cosmological constant.
We conclude the coefficients of such corrections are enhanced compared to the case without bulk supersymmetry.

\end{abstract}
\date{\today}

\maketitle
\flushbottom
\section{Introduction }
\label{sec:intro}
The extra dimension has been discussed for more than 80 years.
Famously, Kaluza and Klein~\cite{Kaluza:1921tu,Klein:1926tv}
introduced an additional compactified dimension and 
suggested that our spacetime needs to have more than four dimensions
to unify gravity and classical electrodynamics.
Now, there have been many extra dimensional models and theories to 
consider various problems of the particle physics.
Especially, string theory~\cite{Horava:1995qa} is a
strong candidate to promise quantum gravity through seven 
additional spatial dimensions.
In string theory (or brane-world scenario) 
we assume that the compactification scale of the extra dimensions is around 
$M_{\rm Planck}\sim 10^{19}{\rm GeV}$. 
Naively, experimental observations of such extra dimensions
are not realistic and we have no choice to extract constraints indirectly.

However, the hierarchy or naturalness 
problem~\cite{Wilson:1970ag,Gildener:1976ai,Weinberg:1978ym,tHooft:1979rat}
strongly circumscribes the possibility of the extra dimensional models.
As the most troublesome issues, 
there have existed the fine-tuning of the Higgs boson
mass and the cosmological constant~\cite{Weinberg:1988cp}.
These dimensional parameter receives quantum corrections of
the ultraviolet (UV) cut-off scale $\Lambda_{\rm UV}$~\cite{Giudice:2013nak} 
and they quadratically or quartically diverge
\footnote{
Many new physics models have been considered~\cite{Randall:1999ee,Graham:2015cka,Matsui:2016cls} until now but there are no obvious solutions.}. These divergences should be regularized as the finite Casimir corrections  depending on the first excited KK scale $1/R$ in the extra dimensional models. The Higgs boson mass  or the cosmological constant in the extra dimensional models 
is formally corrected from infinite KK states of bulk fields.  
The Kaluza-Klein regularization 
assumes exchanging loop integrals 
and summation of all the KK states, and then,
regularizes infinite divergences with KK summation using zeta function formula. 
Adopting this method, we can regularize the infinite KK corrections
and clearly obtain finite values which corresponds to  
the Casimir energy from extra dimension.
However, the validity of the method is still under debate
\cite{Kobayashi:2001zh,Contino:2001gz,Ghilencea:2001ug,Kubo:1999ua,Ghilencea:2001bv,Kim:2001gk,Antoniadis:2001cv,Delgado:2001xr,Barbieri:2000vh,Antoniadis:1990rq,Buchbinder:1990aj} and 
the cut-off sensitivity becomes more unclear.

The main purpose of this paper is to evaluate
quantum radiative corrections for the Higgs mass with 
infinite Kaluza-Klein (KK) states 
and clarify the Casimir vacuum energy from extra dimensions.
In order to avoid the above ambiguity, in this paper,
we evaluate one-loop quantum corrections from the extra dimensions
with the KK regularization or the dimensional regularization. Then, 
we clearly show that the regularized finite corrections 
corresponds to the Casimir energy from the extra dimensions.
The regularized corrections from the KK states have the form of $\propto 1/R^2$ 
for the Higgs mass and $\propto 1/R^4$ for the cosmological constant
(where $R$ is the compactification radius).  
The Casimir corrections can be considered as the quantum effects 
of compactified spaces, which is separated from the power-law divergences 
derived from the momentum cut-off 
or the pole for the dimensional regularization.
Finally, we discuss supersymmetric or
non-supersymmetric Casimir vacuum energy from the extra dimensions.


The present paper is organized as follows.
In Section~\ref{sec:casimir} 
we introduce the Casimir energy from quantum zero-point energy with the Dirichlet boundary condition. We see divergences from $d$-dimensional momentum integrals and infinite summations of discretized modes are regularized 
In Section~\ref{sec:casimir/extra} 
we discuss the regularization issues for the Higgs mass and the cosmological constant 
in five dimensional (5D) models with two boundaries and evaluate the 
radiative corrections from all the KK states of a bulk fermion. 
The divergences from the summation of KK states are
regularized by the analytical continuation of the Riemann zeta function.
However, we clearly show that the finite contributions correspond to the 
Casimir energy from extra dimensions.
In Section~\ref{sec:casimir/susy} 
we consider the case with supersymmetry and evaluate the 
Casimir quantum correction for the Higgs mass and the cosmological constant.
We also discuss the Casimir corrections and divergent parts in suspersymmetric extra dimension.
We comment the difference of the treatment of divergent parts in the KK regularization  and the dimensional regularization with SUSY.
Finally, in Section~\ref{sec:summary} we draw the conclusion of this paper.


\section{Casimir energy }
\label{sec:casimir}
In this section we introduce the Casimir energy,
which is formally defined as the quantum zero-point energy
with the boundary conditions. 
The QFT~\cite{Born1926} is formally constructed as an enormously
large collection of the quantum harmonic oscillators.
Thus, the vacuum energy with various quantum fields receives the divergent 
zero-point energy
\begin{align}
{ E }_{ \rm zero }=\frac { \pm 1 }{ 2 }\sum_{\rm spin }\sum_{k}{\omega_k}\longrightarrow \infty,
\end{align}
where $\omega_k= \sqrt{k^2+m^2}$ and $m$ is the masses of the boson and 
fermion fields. In principle the divergent zero-point quantum corrections 
can be renormalized by the bare parameters and one
fixes the finite physical parameters so that they agree with the observations,
although the fine-tuning between the bare parameters and the quantum corrections
would be still serious.
In this sense, the QFT makes no prediction for the physical values of the 
vacuum energy or cosmological constant~\cite{Maggiore:2010wr}
and there exists no consensus about the reality of the zero-point energy 
in the community of the particle physics.

However, the difference of the divergent zero-point energy $\Delta { E }_{ \rm zero }$
has already been recognized to provide the observable effects.
Famously, the Casimir effect~\cite{Casimir:1948dh}
can be described by the zero-point electromagnetic energy
between two parallel conducting plates and 
has been experimentally detected~\cite{Lamoreaux:1996wh}. 
The difference of the divergent zero-point energy 
with the boundary conditions
becomes finite and physical Casimir energy~\cite{Bordag:2001qi},
and the phenomena have indeed provided 
an important hint of this problem.

Now, let us consider a massless scalar field between
two parallel plates to impose the Dirichlet boundary condition,
\begin{align}
\phi \left(z=0 \right) =\phi \left(z=a\right) =0.
\end{align}
This boundary condition discretize the modes $k={n\pi}/{a}$
and the zero-point energy on the condition can be given by
\begin{align}
{ E }_{ \rm zero }=\frac { 1 }{ 2 }\sum_{k}{\omega_k}=\frac { 1 }{ 2 }\sum_{n=0}^{\infty }
{\int { \frac { { d }^{ 2 }k }{ { \left( 2\pi  \right)  }^{ 2 } } \sqrt { { k }^{ 2 }+{ \left( \frac { n\pi  }{ a }  \right)  }^{ 2 } }  } },
\end{align}
which has some divergences.

Thus, let us adopt so-called zeta function regularization method 
using the following mathematical formula, 
\begin{align}
\int _{ 0 }^{ \infty  }{ \frac { dt }{ t } { t }^{ -\alpha }{ e }^{ -zt }=\Gamma \left( -\alpha \right) { z }^{ \alpha } }, \quad 
\int { { d }^{ d }k{ e }^{ -t{ k }^{ 2 } }={ \left( \frac { \pi  }{ t }  \right)  }^{ d/2 } } ,\label{eq: SchPTI}
\end{align}
where $d$ is the complex dimension of the spacetime
and the left expression is called the proper time integral. 
Using these formulae, we can get the following expression,
\begin{align}
{ E }_{ \rm zero }&=\frac { 1 }{ 2 }\sum_{n=0}^{\infty }
{\int { \frac { { d }^{ d}k }{ { \left( 2\pi  \right)  }^{ d } } \int _{ 0 }^{ \infty  }{ \frac { dt }{t\cdot \Gamma \left( -1/2 \right)  } { t }^{ -1/2 }{ e }^{ -t\left( { k }^{ 2 }+{ \left( { n\pi  }/{ a }  \right)  }^{ 2 }\right)}}}} \\
&=-\frac { 1 }{ 4\sqrt { \pi  }  } \frac { 1 }{ { \left( 4\pi  \right)  }^{ d/2 } } 
\sum _{ n=0 }^{ \infty }{ \int _{ 0 }^{ \infty  }{ \frac { dt }{ t } { t }^{ -1/2-d/2 }{ e }^{ -{ t{ n }^{ 2 }{ \pi  }^{ 2 } }/{ { a }^{ 2 } }  } }  } \nonumber
\end{align}
Proceeding the calculation we obtain 
\begin{align}
{ E }_{ \rm zero }
&=-\frac { 1 }{ 4\sqrt { \pi  }  } \frac { 1 }{ { \left( 4\pi  \right)  }^{ d/2 } } { \left( \frac { \pi  }{ a }  \right)  }^{ 1+d }
\Gamma \left( -\frac { d+1 }{ 2 }  \right) \sum _{ n=0 }^{ \infty  }{ { n }^{ d+1 } }  \\
&=-\frac { 1 }{ 4\sqrt { \pi  }  } \frac { 1 }{ { \left( 4\pi  \right)  }^{ d/2 } } { \left( \frac { \pi  }{ a }  \right)  }^{ 1+d }
\Gamma \left( -\frac { d+1 }{ 2 }  \right) \zeta  \left( -d-1 \right) \nonumber
\end{align}
Now we take analytic continuation to remove the divergences and 
get the Riemann zeta function,
\begin{align}
\Gamma \left( \frac { z }{ 2 }  \right) \zeta \left( z \right) { \pi  }^{ -z/2 }=
\Gamma \left( \frac { 1-z }{ 2 }  \right) \zeta \left( 1-z \right) { \pi  }^{ -\left( 1-z \right) /2 }
\end{align}
The quantum zero-point energy on the Dirichlet
boundary condition can be written as follows:
\begin{align}
{ E }_{ \rm Casimir }
&=\lim _{ d\rightarrow 2
 }{ \left\{  -\frac { 1 }{ { 2 }^{ d+2 }{ \pi  }^{ d/2+1 } } \frac { 1 }{ { a }^{ d+1 } } 
\Gamma \left( 1+\frac { d }{ 2 }  \right) \zeta \left( 2+d \right)\right\} }\nonumber \\ 
&=-\frac { { \pi  }^{ 2 } }{ 1440\cdot { a }^{ 3 } } 
\end{align}
where the zero-point divergences are removed by 
the analytic continuation of the zeta function $ \zeta \left( z \right)$.
The finite and negative contribution of the zero-point energy becomes 
so-called Casimir energy and observable as the attractive force 
between two parallel plates at small distances,
\begin{align}
{ F}_{ \rm Casimir   }=-\frac { \partial { E }_{ \rm Casimir  }  }{ \partial a } 
=-\frac{\pi^2}{480\cdot  a^4}
\end{align}
where ${ F}_{ \rm Casimir}$ is a famous Casimir force per unit area.
In the case of the electromagnetic fields between 
two parallel conductive plates, the Casimir force can be given as
${ F}_{ \rm Casimir  }=-{\pi^2}/{(240\cdot  a^4)}$~\cite{Casimir:1948dh} where we count 
the two polarization states of the photon.
The Casimir effect has been confirmed in many experiments~\cite{Mohideen:1998iz,
Roy:1999dx,Bressi:2002fr}
and strongly depend on the size, geometry and topology 
of the given boundaries. 
On the other hand, whether the observations of the Casimir force prove the 
reality of the zero-point energy or not has been still under debate~\cite{Jaffe:2005vp}
because the Casimir force can alternatively be computed without 
invoking the zero-point electromagnetic energy as the standard perturbative methods of QED~\cite{Jaffe:2005vp} like the Lamb shift and the 
van der Waals interactions~\cite{Lifshitz:1956zz,0038-5670-4-2-R01,Schwinger:1977pa}.
The question of whether the zero-point energy exists or not 
is outside the scope of the present paper.
However, the Casimir energy from geometrical conditions 
can be definitely defined and discussed about the corrections to the physical parameters.
From here let us consider Casimir energy from extra dimensions
and discuss its corrections for the Higgs mass and the cosmological constant.

\section{Casimir corrections from extra dimensions}
\label{sec:casimir/extra}
In this section, we discuss one-loop quantum corrections for the Higgs mass and the cosmological constant
from the infinite KK states using the Kaluza-Klein regularization or 
the dimensional regularization.

\subsection{Casimir corrections for the Higgs mass
with Kaluza-Klein regularization}
The Kaluza-Klein (KK) regularization is one of the methods to regularize
the divergences from infinite KK states and assumes the exchange between the loop integrals 
and the summation, and regularizes divergences of all the KK states
using the zeta function formula.

Following literature~\cite{Kobayashi:2001zh}
let us discuss one-loop corrections for the Higgs mass in 5D models 
where the radius of the compactified space is $R$. 
The contributions from bulk fermions to the Higgs mass are 
expressed as the infinite summations of those of the $n$-th KK states;
\begin{equation}
Y^2\sum_{n=-\infty}^{\infty}\int_0^\infty\frac{d^4p}{(2\pi)^4}\frac{1}{p^2+m_n^2},\label{eq:fermionloop}
\end{equation}
where $Y$ is the 4D reduced Yukawa coupling of each fermionic KK state and 
assumed to be universal among different $n$-th KK states. 
The $n$-th KK mass eigenvalue is expressed as $m_n$ and 
we assume that the $n$-th fermionic KK state has the KK mass eigenvalue $m_n=\frac{n}{R}\pi$.

Now, we utilize the following identities for the proper time integrals in the Schwinger representation as in the descriptions of \cite{Kobayashi:2001zh},
\begin{align}
\int_{0}^{\infty}dte^{-At}=\frac{1}{A},\quad
\int d^4pe^{-p^2t}=\frac{\pi^2}{t^2}.
\end{align}
Then, Eq.~(\ref{eq:fermionloop}) is rewritten as
\begin{equation}
\frac{Y^2}{16\pi^2}\sum_{n=-\infty}^{\infty}\int_0^\infty\frac{dt}{t^2}e^{-\pi^2tn^2/R^2}.\label{eq:fermionPTI}
\end{equation}
The above integrals have divergences at $t=0$. Now, we adopt the cut-off regularization method in calculating the loop integrals. We truncate $t$ by $1/\Lambda^2$ at $t=0$. Additionally, we truncate $n$ by the cut-off KK number $l$ at $n=\infty$ (by $n=-l$ at $n=-\infty$). Here, we define the following integral $I_{l, \Lambda}$,
\begin{align}
I_{l, \Lambda}=\sum_{n=-l}^{l}\int_{1/\Lambda^2}^\infty\frac{dt}{t^2}e^{-\pi^2tn^2/R^2}.
\end{align}
Since the summation and the integral are finite in the above expression, we can exchange them safely,
\begin{equation}
I_{l, \Lambda}=\int_{1/\Lambda^2}^\infty dt\sum_{n=-l}^{l}\frac{1}{t^2}e^{-\pi^2tn^2/R^2}.\label{eq:exchangedPTI}
\end{equation}
Next, we can take the limit of $l\rightarrow \infty$ when $l\gg\Lambda R/\pi$ in Eq.~(\ref{eq:exchangedPTI});
\begin{equation}
I_{\infty, \Lambda}=\int_{1/\Lambda^2}^\infty dt\sum_{n=-\infty}^{\infty}\frac{1}{t^2}e^{-\pi^2tn^2/R^2}.\label{eq:InfinitePTI}
\end{equation}
Now, we use the Poisson resummation formula;
\begin{equation}
\sum_{n=-\infty}^{\infty}e^{-\pi n^2x}=\frac{1}{\sqrt{x}}\sum_{w=-\infty}^{\infty}e^{-\frac{\pi w^2}{x}},
\end{equation}
where $w$ is the winding number of the bulk spacetime. 
Then, Eq.~(\ref{eq:InfinitePTI}) becomes
\begin{align}
I_{\infty, \Lambda}&=I_{w=\infty,\Lambda}=\frac{R}{\sqrt{\pi}}\int_{1/\Lambda^2}^{\infty}dtt^{-\frac{5}{2}}\sum_{w=-\infty}^{\infty}e^{-\frac{R^2}{t}w^2}\nonumber\\
&=\frac{R}{\sqrt{\pi}}\int_{1/\Lambda^2}^{\infty}dtt^{-\frac{5}{2}}\left(1+2\sum_{w=1}^{\infty}e^{-\frac{R^2}{t}w^2}\right).
\end{align}
Similarly to the process from Eq.~(\ref{eq:fermionPTI}) to Eq.~(\ref{eq:exchangedPTI}), we again exchange the summation and the integral as
\begin{equation}
I_{w=\infty, \Lambda}=\frac{R}{\sqrt{\pi}}\left(\int_{1/\Lambda^2}^{\infty}dtt^{-\frac{5}{2}}+2\sum_{w=1}^{\infty}\int_{1/\Lambda^2}^{\infty}dtt^{-\frac{5}{2}}e^{-\frac{R^2}{t}w^2}\right). \label{eq:devidedPTI}
\end{equation}
The first term in the above equation is
\begin{equation}
\frac{2R}{3\sqrt{\pi}}\Lambda^3.
\end{equation}
When we take the limit $\Lambda\rightarrow\infty$, the first term diverges obviously. At the same time, the second term in Eq.~(\ref{eq:devidedPTI}) becomes
\begin{align}
&\frac{2R}{\sqrt{\pi}}\sum_{w=1}^{\infty}\int_{0}^{\infty}dtt^{-\frac{5}{2}}e^{-\frac{R^2}{t}w^2}\nonumber\\
=&-\frac{2R}{\sqrt{\pi}}\cdot\frac{1}{R^3}\sum_{w=1}^{\infty}\frac{1}{w^3}\cdot\int_\infty^0 dyy^{\frac{1}{2}}e^{-y}\nonumber\\
=&\frac{2}{\sqrt{\pi}R^2}\sum_{w=1}^{\infty}\frac{1}{w^3}\cdot\int_0^\infty dyy^{\frac{1}{2}}e^{-y}\nonumber\\
=&\frac{2}{\sqrt{\pi}R^2}\zeta(3)\Gamma\left(\frac{3}{2}\right),
\end{align}
where we transformed as $y=R^2w^2/t$. Eventually,
\begin{align}
\frac{Y^2}{16\pi^2}I_{\infty, \infty}&=\infty+\frac{Y^2}{8\pi^\frac{5}{2}R^2}\zeta(3)\Gamma\left(\frac{3}{2}\right).\nn
&=\infty+0.007612114264598318Y^2/R^2.\label{eq: KKhiggs}
\end{align}
 The second finite contribution corresponds to the Casimir energy for the one-loop corrections for the Higgs mass.
At the finite cut-off scale $\Lambda$, the contributions from bulk fermions are expressed as
\begin{equation}
\frac{Y^2}{16\pi^2}I_{\infty,\Lambda}=\frac{Y^2}{24\pi^\frac{5}{2}}R\Lambda^3+\frac{y^2}{8\pi^\frac{5}{2}R^2}\sum_{w=1}^{\infty}\frac{1}{w^3}\int_0^{R^2\Lambda^2w^2}dyy^\frac{1}{2}e^{-y}.
\end{equation}
We can easily check that the second term is finite at every $R>0$ or $\Lambda>0$.

\subsection{Casimir corrections for the Higgs mass
with dimensional regularization}
However, there remain still some doubtful points in the method of the 
KK regularization~\cite{Kobayashi:2001zh,Contino:2001gz,Ghilencea:2001ug,Kubo:1999ua,Ghilencea:2001bv,Kim:2001gk,Antoniadis:2001cv,Delgado:2001xr,Barbieri:2000vh,Antoniadis:1990rq,Buchbinder:1990aj} . 
The important point is the validity of exchanging the loop integrals and the KK summation
because the infinite integrals and sums can not always be exchanged.
For instance, in the discussions of the literature~\cite{Kobayashi:2001zh}, 
the regularized loop integrals $\int ^\Lmd dp$ by the cut-off regularization and the KK summation $\Sigma_{n=-\infty({\rm or}\ 0)}^\infty$ truncated by the finite cut-off $l$ of the KK level $n$ are exchanged. Next, the authors take the limit of $\Lmd, l=\infty$ in
the exchanged form of the corrections as follows:
\begin{align}&\sum_{n}^\infty\int_0^\infty dp \longrightarrow \sum_{n}^l\int_0^\Lmd dp\nn =& \int_0^\Lmd dp  \sum_{n}^l \longrightarrow \int_0^\infty dp\sum_{n}^\infty.
\end{align}
which seems to be correct.
However, there are no proof of the KK regularization
that the regularized values before exchanging and after exchanging are the same.
Therefore, we must carefully check  and confirm the validity of the KK regularization by
calculating the KK corrections with another regularization method.

In order to avoid the above ambiguity of the KK regularization,
we regularize the 4-momentum integrals by the method of dimensional 
regularization before summing all the KK states, following the paper \cite{Gupta:2002qu}.  
Now, we define the proper time integral $I_n$ as follows:
\bea
I_n&\equiv&\int\frac{d^4p}{(2\pi)^4}\frac{1}{p^2+m_n^2}\nn
&=&\int\frac{d^4p}{(2\pi)^4}\frac{1}{\Gamma(1)}\int_0^\infty dtt^{1-1}e^{-(p^2+m_n^2)t}\nn
&=&\int_0^{\infty}  dt \frac{e^{-m_n^2t}}{(4\pi t)^{4/2}}\nn
&\sim&\Gamma\brkt{1-\frac{d}{2}}\frac{\mu^{4-d}}{(4\pi)^{d/2}}(m_n^2)^{-(1-\frac{d}{2})} \ \ \ (d\sim4)\nn
&\sim&\Gamma\brkt{1-\frac{d}{2}}\frac{m_n^2}{(4\pi)^{d/2}}\brkt{\frac{m_n^2}{\mu^2}}^{\frac{d}{2}-2},\ \ \ (d\sim4)
\label{eq:pti}
\eea
where $\mu$ is the renormalization scale and $d$ is complex dimension of the spacetime. 
We continue Eq.~(\ref{eq:pti}) analytically at $d=4$ and 
rewrite this expression with the infinitesimal parameter $\vep=\frac{d}{2}-2$ as
\bea
I_n&=&\Gamma\brkt{\vep-1}\frac{m_n^2}{(4\pi)^{\vep+2}}\brkt{\frac{m_n^2}{\mu^2}}^\vep\nn
&=&\brc{-\frac{1}{\vep}+\gamma-1+(\cdots)\vep+\cO(\vep^2)}\brkt{\frac{m_n}{4\pi}}^2\brkt{4\pi\frac{\mu^2}{m_n^2}}^\vep,\nn
\eea
where $\gamma=0.577...$ is the Euler's constant.
From 
\be
\brkt{4\pi\frac{\mu^2}{m_n^2}}^\vep\sim1+\vep\ln\brkt{4\pi\frac{\mu^2}{m_n^2}}+\cO(\vep^2),
\ee
$I_n$ is rewritten as
\begin{align}
I_n=\brkt{\frac{m_n}{4\pi}}^2\biggr\{-\frac{1}{\vep}-1+\gamma&-\ln\brkt{4\pi\frac{\mu^2}{m_n^2}}\nn
&+({\rm negligible\ terms})\biggl\}.
\end{align}
Therefore,  $I_f$ is given as follows:
\begin{align}
I_f&=Y^2\sum_{n=-\infty}^{\infty}I_n\nn
&=Y^2\sum_{n=-\infty}^\infty\brkt{\frac{m_n}{4\pi}}^2\brc{-\frac{1}{\vep}-1+\gamma+\ln\brkt{\frac{m_n^2}{4\pi\mu^2}}}.\label{eq:ifmn}
\end{align}

Now, we adopt the Riemann zeta function regularization to 
remove divergences of the quantum corrections,
\bea
\sum_{n=-\infty}^\infty n^2&=&0+2\sum_{n=1}^\infty n^2,\nn
&=&2\zeta(-2)\nn
&=&0,\label{eq:zeta-2}
\eea
and
\begin{align}
\sum_{n=-\infty}^\infty n^2\ln n^2&=0+2\sum_{n=1}^\infty n^2\ln n^2\nn
&=4\sum_{n=1}^\infty n^2\ln n\nn
&=-4\zeta'(-2).
\end{align}
where $\zeta'(x)$ is the differential of the zeta function $\zeta(x)$.

Therefore, the KK summed finite correction in Eq.~(\ref{eq:ifmn}) becomes
\bea
I_{\rm f\ Casimir}=-\frac{Y^2}{4R^2}\zeta'\brkt{-2}
=0.007612114264598319\frac{Y^2}{R^2}.\nn
\label{eq:kkkdifmn}
\eea
which corresponds to the Casimir corrections for the Higgs mass 
from the KK states of bulk fermions.
The value of Eq.~(\ref{eq:kkkdifmn})  agrees highly accurately with the
value of the KK regularization (\ref{eq: KKhiggs}).
On the other hand, the pole term in the expression of Eq.~(\ref{eq:ifmn}),
\be
-\frac{Y^2}{16\pi^2R^2}\sum_{n=-\infty}^{\infty}n^2\cdot\frac{1}{\vep},
\ee
\footnote{In Ref~\cite{Ghilencea:2005hm}, a different analysis is made by the dimensional regularization.  In this case, the UV divergences can be isolated instead of a separate
treatment of the KK sum and 4D momentum integral. They can isolate divergences by exchanging sum and integral in 5D dimensional regularization (see Appendix A of~\cite{Ghilencea:2005hm}), and the divergences are cancelled by higher 
dimensional operators.}
must be treated carefully.
If the limit $n\rightarrow\infty$ is faster than the limit $\vep\rightarrow0$, 
the term is exactly zero by the relation of Eq.~(\ref{eq:zeta-2}). 
However, when the limit $\vep\rightarrow0$ is faster 
than the limit $n\rightarrow\infty$, 
the term unavoidably diverges. 
In the latter case, this pole term must be renormalized
by the bare parameter.
Note that vanishing divergences in the above regularization 
has no bearing on the naturalness problem of the Higgs mass.
Famously, the dimensional regularization reduces quadratic divergences,
but the fine-tuning problem still exists and it's just appearance~\cite{Giudice:2013nak}.
The infinite KK states exacerbate the problem rather than 
four dimensional theory although
the above regularization procedure hides the cut-off sensitivity.

\subsection{Casimir corrections for the cosmological constant
with Kaluza-Klein regularization}
We consider the one-loop quantum corrections
for the cosmological constant in 5D models with bulk fermions
and discuss the Casimir energy from extra dimensions.
 
 First, we evaluate with the Kaluza-Klein regularization. The
zero-point vacuum energy 
derived from all the KK states of the bulk fermions is written as 
\bea
\rho_{\rm zero}=\frac{1}{2}\sum_{n=-\infty}^\infty\int\frac{d^3k}{(2\pi)^3}\sqrt{k^2+m_n^2},\label{eq:KKsumedZPE1}
\eea
where $m_n=\pi\frac{n}{R}$ is the mass eigenvalue of the $n$-th level KK state of the fermion. We use (\ref{eq: SchPTI}) again, 
\begin{align}
\rho_{\rm zero}=-\frac{1}{4\sqrt{\pi}}\frac{1}{(4\pi)^{3/2}}\sum_{n=-\infty}^\infty\int_0^\infty dt t^{-3}e^{-\frac{n^2}{R^2}\pi^2t}.
\end{align}
Then, we permit the exchange of the infinite sum and the divergent integral,
\begin{align}
\rho_{\rm zero}=-\frac{1}{32\pi^2}\int_0^\infty dt t^{-3}\sum_{n=-\infty}^\infty e^{-\frac{n^2}{R^2}\pi^2 t}. 
\end{align}
We use the Poissson resummation formula, 
\begin{align}
\rho_{\rm zero}&=-\frac{1}{32\pi^2}\int_0^\infty dt t^{-3}\cdot\frac{R}{\sqrt{\pi t}}\brkt{1+2\sum_{w=1}^\infty e^{-\frac{R^2}{t}w^2}}\nn
&=-\frac{R}{32\pi^{5/2}}\int_0^\infty dt t^{-7/2} \brkt{1+2\sum_{w=1}^\infty e^{-\frac{R^2}{t}w^2}}.
\end{align}
Now, we truncate $t$ by the finite cut-off $t=1/\Lmd^2$ at $t=0$.
\begin{align}
\rho_{\rm zero}&=-\frac{R}{32\pi^{5/2}}\brkt{\int_{1/\Lmd^2}^\infty dt t^{-7/2}+2\sum_{w=1}^\infty\int_{1/\Lmd^2}^\infty dt t^{-7/2}e^{-\frac{R^2}{t}w^2}}\nn
&=-\frac{R\Lmd^5}{112\pi^{5/2}}-\frac{R}{16\pi^{5/2}}\sum_{w=1}^\infty\int_{1/\Lmd^2}^\infty dt t^{-7/2}e^{-\frac{R^2}{t}w^2}.
\end{align}
When we take the limit $\Lmd\rightarrow\infty$,
\begin{align}
\rho_{\rm zero}&=-\infty-\frac{1}{16\pi^{5/2}R^4}\sum_{w=1}^\infty\frac{1}{w^5}\int_0^\infty dyy^{3/2}e^{-y}\nn
&=-\infty-\frac{1}{16\pi^{5/2}R^4}\zeta(5)\Gamma\brkt{\frac{5}{2}}.
\end{align}
The second finite term is the Casimir correction for the vacuum energy and  written as 
\be
\therefore\ \rho_{\rm Casimir}=-0.0049248162891899275/R^4.\label{eq: KKvacuum}
\ee
\subsection{Casimir corrections for the cosmological constant
with dimensional regularization}

Second, we evaluate the one-loop quantum corrections
for the cosmological constant in 5D models with the dimensional regularization.

Let us consider the case of the four dimension for simplicity.
Adopting the dimensional regularization,
the quantum corrections of the cosmological constant from 
the zero-point energy are given by 
\begin{align}
\rho_{\rm zero}&=\frac{{ E }_{ \rm zero }}{\rm Volume }
=\frac{1}{2}\int { \frac { { d }^{ 3 }k }{ { \left( 2\pi  \right)  }^{ 3 } }
 \sqrt { { k }^{ 2 }+{ m }^{ 2 } }  } \nn
 &= \frac { { m }^{ 4 } }{ 64{ \pi  }^{ 2 } } \left\{ \ln { \left( \frac { { m }^{ 2 } }{ { \mu  }^{ 2 } }  \right)  } 
-\frac { 1 }{ \vep  } -\log { 4{ \pi  } } +\gamma -\frac { 3 }{ 2 }  \right\} 
\end{align}
Then, we divide the bare cosmological constant vacuum term ${ \rho  }_{ \Lambda  }$
to be ${ \rho  }_{ \Lambda  } ={ \rho  }_{ \Lambda  }\left(\mu\right)+\delta{ \rho  }_{ \Lambda  }$
where ${ \rho  }_{ \Lambda  }= \Lambda /{8\pi { G }_{ N }}$ is defined by 
the cosmological constant $\Lambda$ and the Newton's constant ${ G }_{ N }$.
The counterterm $\delta{ \rho  }_{ \Lambda  }$ is written as
\begin{align}
\delta{ \rho  }_{ \Lambda  }=\frac { { m }^{ 4 } }{ 4{ \left( 4\pi  \right)  }^{ 2 } } \left(
\frac { 1 }{ \vep  } +\log { 4{ \pi  } } -\gamma  \right).
\end{align}
where we adopt the $\overline {\rm MS }$ scheme.
Absorbing divergences into the counterterm $\delta{ \rho  }_{ \Lambda  }$, 
we obtain the following renormalized expression,
\begin{align}
{ \rho  }_{\rm vacuum}&={ \rho  }_{ \Lambda  }\left(\mu\right)+\delta{ \rho  }_{ \Lambda  }\nn
&+ \frac { { m }^{ 4 } }{ 64{ \pi  }^{ 2 } } \left\{ \ln { \left( \frac { { m }^{ 2 } }{ { \mu  }^{ 2 } }  \right)  } 
-\frac { 1 }{ \vep  } -\log { 4{ \pi  } } +\gamma -\frac { 3 }{ 2 }  \right\}  \nn
&={ \rho  }_{ \Lambda  }\left(\mu\right)+\frac { { m }^{ 4 }}{ 64{ \pi  }^{ 2 } }
 \left(  \ln {\frac { { m }^{ 2 } }{ { \mu  }^{ 2 } }  }-\frac{3}{2} \right) \label{eq:dsldkfgsdg},
\end{align}
where the divergences of the zero-point vacuum energy are definitely 
renormalized by the cosmological constant term.
In the Standard Model (SM) framework, 
the vacuum energy density can be written as follows:
\begin{align}
&{ \rho  }_{\rm vacuum  }={ \rho  }_{ \Lambda  }\left(\mu\right)
+{ \rho  }_{ \rm  EW }+{ \rho  }_{ \rm  QCD } \nn
&+\sum _{ i }^{  }{  \frac { n_{i}{ m }_{i}^{ 4 }}{ 64{ \pi  }^{ 2 } }
\left( \ln {  \frac { { m }_{i}^{ 4 }  }{ { \mu  }^{ 2 } } } -\frac{3}{2} \right)}+
\mathcal{O}({ \Lambda  }_{ \rm  UV}^4)+\cdots,
\end{align}
where ${ \rho }_{ \rm EW }$ or ${ \rho }_{ \rm QCD }$ express the 
classical vacuum energies of the electroweak or chiral symmetry breaking.
$n_{i}$ and $m_{i}$ are the number of degrees of freedom and 
the mass of the SM particle $i$, respectively. 
However, the dark energy~\cite{Perlmutter:1997zf,Riess:1998cb,Bahcall:1999xn} 
as the current physical value of the vacuum energy
is extremely small. The fine-tuning of the vacuum energy is highly problematic 
and we have no satisfactory solutions to
derive such an extremely small scale.

Next, we evaluate the quantum corrections for the cosmological constant 
in 5D models with bulk fermions.
The 
zero-point vacuum energy 
derived from all the KK states of the bulk fermions is written as 
\bea
\rho_{\rm zero}=\frac{1}{2}\sum_{n=-\infty}^\infty\int\frac{d^3k}{(2\pi)^3}\sqrt{k^2+m_n^2},\label{eq:KKsumedZPE2}
\eea
where $m_n=\pi\frac{n}{R}$ is the mass eigenvalue of the $n$-th level KK state of the fermion.
Then, we rewrite the renormalized expression of Eq.~(\ref{eq:dsldkfgsdg}) as
\begin{align}
{ \rho  }_{\rm vacuum}&=\rho_\Lmd(\mu)+\sum_{n=-\infty}^{\infty}\frac { { m_n }^{ 4 }}{ 64{ \pi  }^{ 2 } }
 \left(  \ln {\frac { { m_n }^{ 2 } }{ { \mu  }^{ 2 } }  }-\frac{3}{2} \right)\nn
 &=\rho_\Lmd(\mu)+\sum_{n=-\infty}^{\infty}\frac{\pi^2 n^4}{64 R^4}
 \brc{\ln n^2+\ln \brkt{\frac{\pi^2}{\mu^2R^2}}-\frac{3}{2}}.\label{eq:rKKsummedZPE}
\end{align}
The KK summed term can be evaluated by 
the Riemann zeta-function regularization:
\be
\sum_{n=-\infty}^{\infty}n^4=0+2\sum_{n=1}^{\infty}\frac{1}{n^{-4}}=2\zeta(-4)=0,
\ee
and
\begin{align}
\sum_{n=-\infty}^{\infty}n^4\ln n^2&=0+2\sum_{n=1}^\infty n^4 \ln n^2\nn
&=4\sum_{n=1}^\infty\frac{1}{n^{-4}}\ln n\nn
&=-4\zeta'(-4),
\end{align}
where $\zeta'(x)$ is the differential of the zeta function $\zeta(x)$ $(\zeta'(-4)$=0.00798381). 
Thus, the KK summed finite contribution in Eq.~(\ref{eq:rKKsummedZPE}) becomes
\begin{align}
{ \rho  }_{\rm Casimir}&=\frac{\pi^2}{64 R^4}\cdot\brc{-4\zeta'(-4)}\nn
&=-0.004924816289189928/R^4
\end{align}
which corresponds to the physical Casimir correction from the extra dimension.
The result agrees very precisely (but  not completely  equal) with the KK regularization on the value of the correction (\ref{eq: KKvacuum}).

\section{Casimir corrections from supersymmetric extra dimensions}
\label{sec:casimir/susy}
Famously,  supersymmetry (SUSY) can in principle remove
quadratic or quartic divergences of the quantum radiative corrections.
However, our real world must break the SUSY and the Higgs mass 
receives quadratic divergent corrections up to the breaking scale. 
The zero-point energy completely cancel out in the SUSY case
due to its opposite signs of the boson and fermion.
However, supersymmetric Casimir energy is still non-zero~\cite{Abe:1999ts,Ghilencea:2005hm,Ghilencea:2004sq,Ghilencea:2005vm}
because the boundary conditions break the SUSY.
As previously discussed in Section~\ref{sec:casimir/extra}, 
the quantum corrections from
extra dimensions correspond to the Casimir energy,
and therefore, the SUSY can not reduce the Casimir quantum corrections 
from the infinite KK states of bulk fields.

\subsection{Supersymmetric Casimir corrections for the Higgs mass
with Kaluza-Klein regularization\label{ss:SUSYHiggsKK}}
Next, we consider bulk supersymmetry and 
discuss bosonic contributions additionally. 
The bosonic contributions are written as~\cite{Kobayashi:2001zh}
\begin{equation}
g^2\int\frac{d^4p}{(2\pi)^4}\frac{1}{p^2+{m'}_n^2},
\end{equation}
where $g$ is the 4D reduced coupling of the bulk SUSY multiplet and the Higgs, and 
it is assumed to be universal among different KK bosonic and fermionic states. 
The bosonic KK mass eigenvalues are expressed as $m'_n$.
Now, we impose the boundary conditions for the bosonic KK modes to have the SUSY breaking KK mass eigenvalues $m'_n=\frac{\pi}{R}(n+\frac{1}{2})$. 
Similarly to Eq.~(\ref{eq:fermionPTI}), 
\begin{align}
&g^2\sum_{n=-\infty}^{\infty}\int_0^\infty\frac{dp^4}{(2\pi^4)}\frac{1}{p^2+{m'}_n^2}\nonumber\\
=&\frac{g^2}{16\pi^2}\sum_{n=-\infty}^{\infty}\int_0^\infty\frac{dt}{t^2}e^{-\pi^2t\left(n+\frac{1}{2}\right)^2/R^2}.
\end{align}
We define the following integral
\begin{equation}
I_{l, \Lambda}^b=\sum_{n=-l}^{l}\int_{1/\Lambda^2}^\infty\frac{dt}{t^2}e^{-\pi^2t\left(n+\frac{1}{2}\right)^2/R^2}.
\end{equation}
When we take the limit $l\rightarrow\infty$,
\begin{equation}
I_{\infty, \Lambda}^b=\int_{1/\Lambda^2}^\infty dt\sum_{n=-\infty}^{\infty}\frac{1}{t^2}e^{-\pi^2t\left(n+\frac{1}{2}\right)^2/R^2}.\label{eq:exchengedBPTI}
\end{equation}
The Poisson resummation formula is rewritten as
\begin{align}
\sum_{n=-\infty}^{\infty}e^{-\pi\left(n+\frac{1}{2}\right)^2x}&=\frac{1}{\sqrt{x}}\sum_{w=-\infty}^\infty e^{\pi iw-\pi\frac{w^2}{x}}\nonumber\\
&=\frac{1}{\sqrt{x}}\sum_{w=-\infty}^\infty (-1)^we^{-\pi\frac{w^2}{x}}.
\end{align}
Then, Eq.~(\ref{eq:exchengedBPTI}) becomes
\begin{align}
I_{\infty, \Lambda}^b&=I_{w=\infty,\Lambda}^b\nonumber\\
&=\frac{R}{\sqrt{\pi}}\int_{1/\Lambda^2}^{\infty}dtt^{-\frac{5}{2}}\sum_{w=-\infty}^{\infty}(-1)^we^{-\frac{R^2}{t}w^2}\nonumber\\
&=\frac{R}{\sqrt{\pi}}\int_{1/\Lambda^2}^{\infty}dtt^{-\frac{5}{2}}\left(1+2\sum_{w=1}^{\infty}(-1)^we^{-\frac{R^2}{t}w^2}\right).
\end{align}

When we take the limit $\Lambda\rightarrow\infty$ and the fermionic contributions are written as
\begin{equation}
g^2\sum_{n=-\infty}^{\infty}\int_0^\infty\frac{dp^4}{(2\pi)^4}\frac{1}{p^2+\left(\frac{n}{R}\pi\right)^2},
\end{equation}
the finite Casimir correction parts in the contributions from both fermionic and bosonic modes are summed up as
\begin{align}
  &\frac{g^2}{16\pi^2}\cdot\frac{2}{\sqrt{\pi}R^2}\sum_{w=1}^{\infty}\frac{1-(-1)^w}{w^3}\cdot\int_0^\infty dyy^\frac{1}{2}e^{-y}\nonumber\\
=&\frac{g^2}{8\pi^\frac{5}{2}R^2}\cdot\frac{7}{4}\zeta(3)\Gamma\left(\frac{3}{2}\right)\nonumber\\
=&
0.013321199963047056g^2/R^2.\label{eq: SUSYKKHiggs}
\end{align}
The power law contributions ($\propto R\Lambda^3$) are exactly offset by the bulk supersymmetry, which is broken softly by the boundary conditions. Because of the boundary conditions, the Casimir energy appears and its coefficient is enhanced compared to the case with only fermionic modes. 
As later discussed in Section~\ref{sec:casimir/susy}, 
this fact that the Casimir energy is enhanced with the SUSY 
is seen in the case of the dimensional regularization.

\subsection{Supersymmetric Casimir corrections for the Higgs mass with the dimensional regularization}

In this section, we evaluate one-loop corrections for the Higgs mass 
 from fermionic and bosonic KK modes in 
5D supersymmetric models and clearly discuss the 
supersymmetric Casimir corrections from the extra dimensions.
We choose boundary conditions that bulk bosons have SUSY breaking KK mass eigenvalues, and  one-loop contributions from all the KK states of the bulk boson are written as
\begin{equation}
g^2\int\frac{d^4p}{(2\pi)^4}\frac{1}{p^2+{m'}_n^2},
\end{equation}
where $m'_n=\frac{\pi}{R}(n+\frac{1}{2})$. 
This is calculated by substituting $n+\frac{1}{2}$ for $n$ in $I_f$. 
To do so, we must evaluate the following quantities;
\begin{align}
\sum_{n=-\infty}^{\infty}\brkt{n+\frac{1}{2}}^2\ \ \&\ \ \sum_{n=-\infty}^{\infty}\brkt{n+\frac{1}{2}}^2\ln\brkt{n+\frac{1}{2}}^2.\label{eq:inftysum}
\end{align}
The first quantity can be exprressed as 0 by Hurvitz zeta function $\zeta(-2, \frac{1}{2})=0$. The second quantity is
\begin{align}
&\sum_{n=-\infty}^{-1}\brkt{n+\frac{1}{2}}^2\ln\brkt{n+\frac{1}{2}}^2\nn
+&\sum_{n=0}^{\infty}\brkt{n+\frac{1}{2}}^2\ln\brkt{n+\frac{1}{2}}^2.
\end{align}
The first term in this expression is
\begin{align}
\sum_{n=1}^\infty\brkt{n-\frac{1}{2}}^2\ln\brkt{n-\frac{1}{2}}^2&=\frac{1}{4}\ln\frac{1}{4}+\frac{9}{4}\ln\frac{9}{4}+\cdots.
\end{align}
This is equal to the second term that is $2\zeta'(-2,\frac{1}{2})$ (the differential of Hurvitz zeta function). Therefore, the second quantity in Eq.~(\ref{eq:inftysum}) is $4\zeta'(-2, \frac{1}{2})$. 
So the KK summed finite parts in the bosonic contributions are
\be
I_b|_{\rm KK}=\frac{g^2}{(4R)^2}\cdot4\zeta'\brkt{-2, \frac{1}{2}}=\frac{g^2}{4R^2}\zeta'\brkt{-2, \frac{1}{2}}.
\ee
As we see, the finite Casimir correction parts derived from the contributions of all the KK states are written as
\bea
I_{\rm f}-I_{\rm b}
&=&I_f|_{\rm KK}-I_b|_{\rm KK}\nn
&=&-\frac{g^2}{4R^2}\brc{\zeta'\brkt{-2}-\zeta'\brkt{-2,\frac{1}{2}}}\nn
&=&0.013321199963047058g^2/R^2
\eea
The result agrees highly accurately with the value of the KK regularization (\ref{eq: SUSYKKHiggs}).

\subsection{Supersymmetric Casimir corrections for the cosmological constant with the Kaluza-Klein regularization}
Similarly, the Casimir
vacuum energy in 5D supersymmetric models can be evaluated.  First, we calculate with the Kaluza-Klein regularization.
We assume the KK mass eigenvalues are the same as the subsection \ref{ss:SUSYHiggsKK}. The Casimir vacuum energy is evaluated as the difference of fermionic  and bosonic contributions;
\begin{align}
\rho_{\rm Casimir}&=-\frac{1}{16\pi^{5/2}R^4}\sum_{w=1}^\infty\frac{1-(-1)^w}{w^5}\int_0^\infty dyy^{3/2}e^{-y}\nn
&=-\frac{1}{16\pi^{5/2}R^4}\cdot\frac{31}{16}\zeta(5)\Gamma\brkt{\frac{5}{2}}\nn
&=-0.009541831560305483/R^4.\label{eq: SUSYKKVE}
\end{align}
We can see the coefficient of the Casimir vacuum energy is enhanced as same as is the case with the Higgs mass.

\subsection{Supersymmetric Casimir corrections for the cosmological constant with the dimensional regularization}
Next, we evaluate the Casimir vacuum energy with the dimensional regularization.
Fermionic and bosonic contributions are calculated by Eq.~(\ref{eq:rKKsummedZPE}). 
The KK summed corrections are regularized by the zeta function regularization. Eventually, the Casimir vacuum energy is evaluated as
\begin{align}
{ \rho  }_{\rm Casimir}&=\frac{\pi^2}{64 R^4}\cdot\sbk{-4\brc{\zeta'(-4)-\zeta'\brkt{-4,\frac{1}{2}}}}\nn
&=-0.009541831560305485/R^4.
\end{align}
The result agrees highly accurately with the value of the KK regularization (\ref{eq: SUSYKKVE}).

In both the case of the Higgs mass and the vacuum energy, the infinite summations of the products of $m_n^2$ or $m_n^4$ and the poles $1/\vep$ diverge when we regularize by the method of the dimensional regularization if the limit $\vep\rightarrow0$ is earlier than the limit of $n\rightarrow\infty$ even with supersymmetry. 
In the cut-off regularization,  the corrections for the Casimir vacuum energy with SUSY is written as
\begin{align}
&\rho_{\rm zero}= \frac{1}{16R^2}\sum_{n=-\infty}^{\infty}
\brc{n^2-\brkt{n+\frac{1}{2}}^2}\Lmd_{ \rm  UV}^2\nn
&+\frac{\pi^2}{64R^4}\sum_{n=-\infty}^\infty \brc{n^4\ln\brkt{\frac{n^2}{\Lmd_{ \rm  UV}^2R^2}}-\brkt{n+\frac{1}{2}}^4\ln\brkt{\frac{n+\frac{1}{2}}{\Lmd_{ \rm  UV}R}}^2}\nn&+\cdots.
\end{align}
In this expression, the contributions of $\Lmd^4$ do not appear because SUSY is broken softly by the boundary conditions.

\section{Summary and conclusion }
\label{sec:summary}

In the present paper, we have reinvestigated regularization of the one-loop quantum corrections for the Higgs mass and the cosmological constant in 5D spacetime
with infinite KK states and discussed the Casimir corrections from extra dimensions.
We evaluated the corrections by the KK regularization and the dimensional regularization.
We found that the Casimir corrections of the KK regularization and the dimensional regularization match with very high accuracy, but  are not completely the same values. These subtle differences may be caused by whether the infinite momentum integrals and the infinite KK summations are exchanged or not. 

We have also evaluated the Casimir corrections in supersymmetric extra-dimensional models. 
The contributions from bulk fermions and bulk bosons (components of bulk SUSY multiplets) 
are not offset because the general boundary conditions break SUSY.
 The supersymmetric Casimir corrections from extra dimensions are still non-zero 
and undoubtedly contributes to the Higgs mass and the cosmological constant.

We have also got the result that the coefficients  of the finite Casimir corrections are 
enhanced a little
with bulk supersymmetry (broken by boundary conditions) compared to the case with only fermionic modes in the calculations of both the Higgs and the vacuum energy. This fact is
seen when we calculate both with the KK regularization and the dimensional regularization method.

Furthermore, we have also discussed the regularization issues of the extra dimension. 
In the dimensional regularization, the infinite KK summations of the products of $n^2$ or $n^4$ and poles $1/\vep$ have divergences that must be renormalized away by the counter terms unlike in the case of the cut-off regularization where the power law contributions of cut-off $\Lmd$ are offset exactly even with SUSY breaking boundary conditions.

\acknowledgments
We thank Yoshiyuki Tatsuta for helpful comments and discussions.

\bibliography{Kaluza-Klein}

\begin{thebibliography}{43}%
\makeatletter
\providecommand \@ifxundefined [1]{%
 \@ifx{#1\undefined}
}%
\providecommand \@ifnum [1]{%
 \ifnum #1\expandafter \@firstoftwo
 \else \expandafter \@secondoftwo
 \fi
}%
\providecommand \@ifx [1]{%
 \ifx #1\expandafter \@firstoftwo
 \else \expandafter \@secondoftwo
 \fi
}%
\providecommand \natexlab [1]{#1}%
\providecommand \enquote  [1]{``#1''}%
\providecommand \bibnamefont  [1]{#1}%
\providecommand \bibfnamefont [1]{#1}%
\providecommand \citenamefont [1]{#1}%
\providecommand \href@noop [0]{\@secondoftwo}%
\providecommand \href [0]{\begingroup \@sanitize@url \@href}%
\providecommand \@href[1]{\@@startlink{#1}\@@href}%
\providecommand \@@href[1]{\endgroup#1\@@endlink}%
\providecommand \@sanitize@url [0]{\catcode `\\12\catcode `\$12\catcode
  `\&12\catcode `\#12\catcode `\^12\catcode `\_12\catcode `\%12\relax}%
\providecommand \@@startlink[1]{}%
\providecommand \@@endlink[0]{}%
\providecommand \url  [0]{\begingroup\@sanitize@url \@url }%
\providecommand \@url [1]{\endgroup\@href {#1}{\urlprefix }}%
\providecommand \urlprefix  [0]{URL }%
\providecommand \Eprint [0]{\href }%
\providecommand \doibase [0]{http://dx.doi.org/}%
\providecommand \selectlanguage [0]{\@gobble}%
\providecommand \bibinfo  [0]{\@secondoftwo}%
\providecommand \bibfield  [0]{\@secondoftwo}%
\providecommand \translation [1]{[#1]}%
\providecommand \BibitemOpen [0]{}%
\providecommand \bibitemStop [0]{}%
\providecommand \bibitemNoStop [0]{.\EOS\space}%
\providecommand \EOS [0]{\spacefactor3000\relax}%
\providecommand \BibitemShut  [1]{\csname bibitem#1\endcsname}%
\let\auto@bib@innerbib\@empty
\bibitem [{\citenamefont {Kaluza}(1921)}]{Kaluza:1921tu}%
  \BibitemOpen
  \bibfield  {author} {\bibinfo {author} {\bibfnamefont {T.}~\bibnamefont
  {Kaluza}},\ }\bibfield  {booktitle} {\emph {\bibinfo {booktitle}
  {{International School of Cosmology and Gravitation: 8th Course: Unified
  Field Theories of More than Four Dimensions, Including Exact Solutions Erice,
  Italy, May 20-June 1, 1982}}},\ }\href@noop {} {\bibfield  {journal}
  {\bibinfo  {journal} {Sitzungsber. Preuss. Akad. Wiss. Berlin (Math. Phys.)}\
  }\textbf {\bibinfo {volume} {1921}},\ \bibinfo {pages} {966} (\bibinfo {year}
  {1921})},\ \Eprint {http://arxiv.org/abs/1803.08616} {arXiv:1803.08616
  [physics.hist-ph]} \BibitemShut {NoStop}%
\bibitem [{\citenamefont {Klein}(1926)}]{Klein:1926tv}%
  \BibitemOpen
  \bibfield  {author} {\bibinfo {author} {\bibfnamefont {O.}~\bibnamefont
  {Klein}},\ }\bibfield  {booktitle} {\emph {\bibinfo {booktitle} {{Gauge
  theories in the twentieth century}}},\ }\href {\doibase 10.1007/BF01397481}
  {\bibfield  {journal} {\bibinfo  {journal} {Z. Phys.}\ }\textbf {\bibinfo
  {volume} {37}},\ \bibinfo {pages} {895} (\bibinfo {year} {1926})},\ \bibinfo
  {note} {[,76(1926)]}\BibitemShut {NoStop}%
\bibitem [{\citenamefont {Horava}\ and\ \citenamefont
  {Witten}(1996)}]{Horava:1995qa}%
  \BibitemOpen
  \bibfield  {author} {\bibinfo {author} {\bibfnamefont {P.}~\bibnamefont
  {Horava}}\ and\ \bibinfo {author} {\bibfnamefont {E.}~\bibnamefont
  {Witten}},\ }\href {\doibase 10.1016/0550-3213(95)00621-4} {\bibfield
  {journal} {\bibinfo  {journal} {Nucl. Phys.}\ }\textbf {\bibinfo {volume}
  {B460}},\ \bibinfo {pages} {506} (\bibinfo {year} {1996})},\ \bibinfo {note}
  {[,397(1995)]},\ \Eprint {http://arxiv.org/abs/hep-th/9510209}
  {arXiv:hep-th/9510209 [hep-th]} \BibitemShut {NoStop}%
\bibitem [{\citenamefont {Wilson}(1971)}]{Wilson:1970ag}%
  \BibitemOpen
  \bibfield  {author} {\bibinfo {author} {\bibfnamefont {K.~G.}\ \bibnamefont
  {Wilson}},\ }\href {\doibase 10.1103/PhysRevD.3.1818} {\bibfield  {journal}
  {\bibinfo  {journal} {Phys. Rev.}\ }\textbf {\bibinfo {volume} {D3}},\
  \bibinfo {pages} {1818} (\bibinfo {year} {1971})}\BibitemShut {NoStop}%
\bibitem [{\citenamefont {Gildener}(1976)}]{Gildener:1976ai}%
  \BibitemOpen
  \bibfield  {author} {\bibinfo {author} {\bibfnamefont {E.}~\bibnamefont
  {Gildener}},\ }\href {\doibase 10.1103/PhysRevD.14.1667} {\bibfield
  {journal} {\bibinfo  {journal} {Phys. Rev.}\ }\textbf {\bibinfo {volume}
  {D14}},\ \bibinfo {pages} {1667} (\bibinfo {year} {1976})}\BibitemShut
  {NoStop}%
\bibitem [{\citenamefont {Weinberg}(1979)}]{Weinberg:1978ym}%
  \BibitemOpen
  \bibfield  {author} {\bibinfo {author} {\bibfnamefont {S.}~\bibnamefont
  {Weinberg}},\ }\href {\doibase 10.1016/0370-2693(79)90248-X} {\bibfield
  {journal} {\bibinfo  {journal} {Phys. Lett.}\ }\textbf {\bibinfo {volume}
  {82B}},\ \bibinfo {pages} {387} (\bibinfo {year} {1979})}\BibitemShut
  {NoStop}%
\bibitem [{\citenamefont {'t~Hooft}(1980)}]{tHooft:1979rat}%
  \BibitemOpen
  \bibfield  {author} {\bibinfo {author} {\bibfnamefont {G.}~\bibnamefont
  {'t~Hooft}},\ }\bibfield  {booktitle} {\emph {\bibinfo {booktitle} {{Recent
  Developments in Gauge Theories. Proceedings, Nato Advanced Study Institute,
  Cargese, France, August 26 - September 8, 1979}}},\ }\href {\doibase
  10.1007/978-1-4684-7571-5_9} {\bibfield  {journal} {\bibinfo  {journal} {NATO
  Sci. Ser. B}\ }\textbf {\bibinfo {volume} {59}},\ \bibinfo {pages} {135}
  (\bibinfo {year} {1980})}\BibitemShut {NoStop}%
\bibitem [{\citenamefont {Weinberg}(1989)}]{Weinberg:1988cp}%
  \BibitemOpen
  \bibfield  {author} {\bibinfo {author} {\bibfnamefont {S.}~\bibnamefont
  {Weinberg}},\ }\href {\doibase 10.1103/RevModPhys.61.1} {\bibfield  {journal}
  {\bibinfo  {journal} {Rev. Mod. Phys.}\ }\textbf {\bibinfo {volume} {61}},\
  \bibinfo {pages} {1} (\bibinfo {year} {1989})},\ \bibinfo {note}
  {[,569(1988)]}\BibitemShut {NoStop}%
\bibitem [{\citenamefont {Giudice}(2013)}]{Giudice:2013nak}%
  \BibitemOpen
  \bibfield  {author} {\bibinfo {author} {\bibfnamefont {G.~F.}\ \bibnamefont
  {Giudice}},\ }\bibfield  {booktitle} {\emph {\bibinfo {booktitle}
  {{Proceedings, 2013 European Physical Society Conference on High Energy
  Physics (EPS-HEP 2013): Stockholm, Sweden, July 18-24, 2013}}},\ }\href@noop
  {} {\bibfield  {journal} {\bibinfo  {journal} {PoS}\ }\textbf {\bibinfo
  {volume} {EPS-HEP2013}},\ \bibinfo {pages} {163} (\bibinfo {year} {2013})},\
  \Eprint {http://arxiv.org/abs/1307.7879} {arXiv:1307.7879 [hep-ph]}
  \BibitemShut {NoStop}%
\bibitem [{\citenamefont {Randall}\ and\ \citenamefont
  {Sundrum}(1999)}]{Randall:1999ee}%
  \BibitemOpen
  \bibfield  {author} {\bibinfo {author} {\bibfnamefont {L.}~\bibnamefont
  {Randall}}\ and\ \bibinfo {author} {\bibfnamefont {R.}~\bibnamefont
  {Sundrum}},\ }\href {\doibase 10.1103/PhysRevLett.83.3370} {\bibfield
  {journal} {\bibinfo  {journal} {Phys. Rev. Lett.}\ }\textbf {\bibinfo
  {volume} {83}},\ \bibinfo {pages} {3370} (\bibinfo {year} {1999})},\ \Eprint
  {http://arxiv.org/abs/hep-ph/9905221} {arXiv:hep-ph/9905221 [hep-ph]}
  \BibitemShut {NoStop}%
\bibitem [{\citenamefont {Graham}\ \emph {et~al.}(2015)\citenamefont {Graham},
  \citenamefont {Kaplan},\ and\ \citenamefont {Rajendran}}]{Graham:2015cka}%
  \BibitemOpen
  \bibfield  {author} {\bibinfo {author} {\bibfnamefont {P.~W.}\ \bibnamefont
  {Graham}}, \bibinfo {author} {\bibfnamefont {D.~E.}\ \bibnamefont {Kaplan}},
  \ and\ \bibinfo {author} {\bibfnamefont {S.}~\bibnamefont {Rajendran}},\
  }\href {\doibase 10.1103/PhysRevLett.115.221801} {\bibfield  {journal}
  {\bibinfo  {journal} {Phys. Rev. Lett.}\ }\textbf {\bibinfo {volume} {115}},\
  \bibinfo {pages} {221801} (\bibinfo {year} {2015})},\ \Eprint
  {http://arxiv.org/abs/1504.07551} {arXiv:1504.07551 [hep-ph]} \BibitemShut
  {NoStop}%
\bibitem [{\citenamefont {Matsui}\ and\ \citenamefont
  {Matsumoto}(2016)}]{Matsui:2016cls}%
  \BibitemOpen
  \bibfield  {author} {\bibinfo {author} {\bibfnamefont {H.}~\bibnamefont
  {Matsui}}\ and\ \bibinfo {author} {\bibfnamefont {Y.}~\bibnamefont
  {Matsumoto}},\ }\href@noop {} {\  (\bibinfo {year} {2016})},\ \Eprint
  {http://arxiv.org/abs/1608.08838} {arXiv:1608.08838 [hep-ph]} \BibitemShut
  {NoStop}%
\bibitem [{\citenamefont {Kobayashi}\ and\ \citenamefont
  {Terao}(2002)}]{Kobayashi:2001zh}%
  \BibitemOpen
  \bibfield  {author} {\bibinfo {author} {\bibfnamefont {T.}~\bibnamefont
  {Kobayashi}}\ and\ \bibinfo {author} {\bibfnamefont {H.}~\bibnamefont
  {Terao}},\ }\href {\doibase 10.1143/PTP.107.785} {\bibfield  {journal}
  {\bibinfo  {journal} {Prog. Theor. Phys.}\ }\textbf {\bibinfo {volume}
  {107}},\ \bibinfo {pages} {785} (\bibinfo {year} {2002})},\ \Eprint
  {http://arxiv.org/abs/hep-ph/0108072} {arXiv:hep-ph/0108072 [hep-ph]}
  \BibitemShut {NoStop}%
\bibitem [{\citenamefont {Contino}\ and\ \citenamefont
  {Pilo}(2001)}]{Contino:2001gz}%
  \BibitemOpen
  \bibfield  {author} {\bibinfo {author} {\bibfnamefont {R.}~\bibnamefont
  {Contino}}\ and\ \bibinfo {author} {\bibfnamefont {L.}~\bibnamefont {Pilo}},\
  }\href {\doibase 10.1016/S0370-2693(01)01352-1} {\bibfield  {journal}
  {\bibinfo  {journal} {Phys. Lett.}\ }\textbf {\bibinfo {volume} {B523}},\
  \bibinfo {pages} {347} (\bibinfo {year} {2001})},\ \Eprint
  {http://arxiv.org/abs/hep-ph/0104130} {arXiv:hep-ph/0104130 [hep-ph]}
  \BibitemShut {NoStop}%
\bibitem [{\citenamefont {Ghilencea}\ and\ \citenamefont
  {Nilles}(2001)}]{Ghilencea:2001ug}%
  \BibitemOpen
  \bibfield  {author} {\bibinfo {author} {\bibfnamefont {D.~M.}\ \bibnamefont
  {Ghilencea}}\ and\ \bibinfo {author} {\bibfnamefont {H.-P.}\ \bibnamefont
  {Nilles}},\ }\href {\doibase 10.1016/S0370-2693(01)00468-3} {\bibfield
  {journal} {\bibinfo  {journal} {Phys. Lett.}\ }\textbf {\bibinfo {volume}
  {B507}},\ \bibinfo {pages} {327} (\bibinfo {year} {2001})},\ \Eprint
  {http://arxiv.org/abs/hep-ph/0103151} {arXiv:hep-ph/0103151 [hep-ph]}
  \BibitemShut {NoStop}%
\bibitem [{\citenamefont {Kubo}\ \emph {et~al.}(2000)\citenamefont {Kubo},
  \citenamefont {Terao},\ and\ \citenamefont {Zoupanos}}]{Kubo:1999ua}%
  \BibitemOpen
  \bibfield  {author} {\bibinfo {author} {\bibfnamefont {J.}~\bibnamefont
  {Kubo}}, \bibinfo {author} {\bibfnamefont {H.}~\bibnamefont {Terao}}, \ and\
  \bibinfo {author} {\bibfnamefont {G.}~\bibnamefont {Zoupanos}},\ }\href
  {\doibase 10.1016/S0550-3213(00)00020-1} {\bibfield  {journal} {\bibinfo
  {journal} {Nucl. Phys.}\ }\textbf {\bibinfo {volume} {B574}},\ \bibinfo
  {pages} {495} (\bibinfo {year} {2000})},\ \Eprint
  {http://arxiv.org/abs/hep-ph/9910277} {arXiv:hep-ph/9910277 [hep-ph]}
  \BibitemShut {NoStop}%
\bibitem [{\citenamefont {Ghilencea}\ \emph {et~al.}(2002)\citenamefont
  {Ghilencea}, \citenamefont {Nilles},\ and\ \citenamefont
  {Stieberger}}]{Ghilencea:2001bv}%
  \BibitemOpen
  \bibfield  {author} {\bibinfo {author} {\bibfnamefont {D.~M.}\ \bibnamefont
  {Ghilencea}}, \bibinfo {author} {\bibfnamefont {H.~P.}\ \bibnamefont
  {Nilles}}, \ and\ \bibinfo {author} {\bibfnamefont {S.}~\bibnamefont
  {Stieberger}},\ }\href {\doibase 10.1088/1367-2630/4/1/315} {\bibfield
  {journal} {\bibinfo  {journal} {New J. Phys.}\ }\textbf {\bibinfo {volume}
  {4}},\ \bibinfo {pages} {15} (\bibinfo {year} {2002})},\ \Eprint
  {http://arxiv.org/abs/hep-th/0108183} {arXiv:hep-th/0108183 [hep-th]}
  \BibitemShut {NoStop}%
\bibitem [{\citenamefont {Kim}(2001)}]{Kim:2001gk}%
  \BibitemOpen
  \bibfield  {author} {\bibinfo {author} {\bibfnamefont {H.~D.}\ \bibnamefont
  {Kim}},\ }\href@noop {} {\  (\bibinfo {year} {2001})},\ \Eprint
  {http://arxiv.org/abs/hep-ph/0106072} {arXiv:hep-ph/0106072 [hep-ph]}
  \BibitemShut {NoStop}%
\bibitem [{\citenamefont {Antoniadis}\ \emph {et~al.}(2001)\citenamefont
  {Antoniadis}, \citenamefont {Benakli},\ and\ \citenamefont
  {Quiros}}]{Antoniadis:2001cv}%
  \BibitemOpen
  \bibfield  {author} {\bibinfo {author} {\bibfnamefont {I.}~\bibnamefont
  {Antoniadis}}, \bibinfo {author} {\bibfnamefont {K.}~\bibnamefont {Benakli}},
  \ and\ \bibinfo {author} {\bibfnamefont {M.}~\bibnamefont {Quiros}},\ }\href
  {\doibase 10.1088/1367-2630/3/1/320} {\bibfield  {journal} {\bibinfo
  {journal} {New J. Phys.}\ }\textbf {\bibinfo {volume} {3}},\ \bibinfo {pages}
  {20} (\bibinfo {year} {2001})},\ \Eprint
  {http://arxiv.org/abs/hep-th/0108005} {arXiv:hep-th/0108005 [hep-th]}
  \BibitemShut {NoStop}%
\bibitem [{\citenamefont {Delgado}\ \emph {et~al.}(2001)\citenamefont
  {Delgado}, \citenamefont {von Gersdorff},\ and\ \citenamefont
  {Quiros}}]{Delgado:2001xr}%
  \BibitemOpen
  \bibfield  {author} {\bibinfo {author} {\bibfnamefont {A.}~\bibnamefont
  {Delgado}}, \bibinfo {author} {\bibfnamefont {G.}~\bibnamefont {von
  Gersdorff}}, \ and\ \bibinfo {author} {\bibfnamefont {M.}~\bibnamefont
  {Quiros}},\ }\href {\doibase 10.1016/S0550-3213(01)00380-7} {\bibfield
  {journal} {\bibinfo  {journal} {Nucl. Phys.}\ }\textbf {\bibinfo {volume}
  {B613}},\ \bibinfo {pages} {49} (\bibinfo {year} {2001})},\ \Eprint
  {http://arxiv.org/abs/hep-ph/0107233} {arXiv:hep-ph/0107233 [hep-ph]}
  \BibitemShut {NoStop}%
\bibitem [{\citenamefont {Barbieri}\ \emph {et~al.}(2001)\citenamefont
  {Barbieri}, \citenamefont {Hall},\ and\ \citenamefont
  {Nomura}}]{Barbieri:2000vh}%
  \BibitemOpen
  \bibfield  {author} {\bibinfo {author} {\bibfnamefont {R.}~\bibnamefont
  {Barbieri}}, \bibinfo {author} {\bibfnamefont {L.~J.}\ \bibnamefont {Hall}},
  \ and\ \bibinfo {author} {\bibfnamefont {Y.}~\bibnamefont {Nomura}},\ }\href
  {\doibase 10.1103/PhysRevD.63.105007} {\bibfield  {journal} {\bibinfo
  {journal} {Phys. Rev.}\ }\textbf {\bibinfo {volume} {D63}},\ \bibinfo {pages}
  {105007} (\bibinfo {year} {2001})},\ \Eprint
  {http://arxiv.org/abs/hep-ph/0011311} {arXiv:hep-ph/0011311 [hep-ph]}
  \BibitemShut {NoStop}%
\bibitem [{\citenamefont {Antoniadis}\ \emph {et~al.}(1990)\citenamefont
  {Antoniadis}, \citenamefont {Ellis}, \citenamefont {Lahanas},\ and\
  \citenamefont {Nanopoulos}}]{Antoniadis:1990rq}%
  \BibitemOpen
  \bibfield  {author} {\bibinfo {author} {\bibfnamefont {I.}~\bibnamefont
  {Antoniadis}}, \bibinfo {author} {\bibfnamefont {J.~R.}\ \bibnamefont
  {Ellis}}, \bibinfo {author} {\bibfnamefont {A.~B.}\ \bibnamefont {Lahanas}},
  \ and\ \bibinfo {author} {\bibfnamefont {D.~V.}\ \bibnamefont {Nanopoulos}},\
  }\href {\doibase 10.1016/0370-2693(90)91480-Y} {\bibfield  {journal}
  {\bibinfo  {journal} {Phys. Lett.}\ }\textbf {\bibinfo {volume} {B241}},\
  \bibinfo {pages} {24} (\bibinfo {year} {1990})}\BibitemShut {NoStop}%
\bibitem [{\citenamefont {Buchbinder}\ \emph {et~al.}(1990)\citenamefont
  {Buchbinder}, \citenamefont {Odintsov},\ and\ \citenamefont
  {Fonarev}}]{Buchbinder:1990aj}%
  \BibitemOpen
  \bibfield  {author} {\bibinfo {author} {\bibfnamefont {I.~L.}\ \bibnamefont
  {Buchbinder}}, \bibinfo {author} {\bibfnamefont {S.~D.}\ \bibnamefont
  {Odintsov}}, \ and\ \bibinfo {author} {\bibfnamefont {O.~A.}\ \bibnamefont
  {Fonarev}},\ }\href {\doibase 10.1016/0370-2693(90)90660-X} {\bibfield
  {journal} {\bibinfo  {journal} {Phys. Lett.}\ }\textbf {\bibinfo {volume}
  {B245}},\ \bibinfo {pages} {365} (\bibinfo {year} {1990})},\ \bibinfo {note}
  {[Pisma Zh. Eksp. Teor. Fiz.51,343(1990)]}\BibitemShut {NoStop}%
\bibitem [{\citenamefont {Born}\ \emph {et~al.}(1926)\citenamefont {Born},
  \citenamefont {Heisenberg},\ and\ \citenamefont {Jordan}}]{Born1926}%
  \BibitemOpen
  \bibfield  {author} {\bibinfo {author} {\bibfnamefont {M.}~\bibnamefont
  {Born}}, \bibinfo {author} {\bibfnamefont {W.}~\bibnamefont {Heisenberg}}, \
  and\ \bibinfo {author} {\bibfnamefont {P.}~\bibnamefont {Jordan}},\ }\href
  {\doibase 10.1007/BF01379806} {\bibfield  {journal} {\bibinfo  {journal}
  {Zeitschrift f{\"u}r Physik}\ }\textbf {\bibinfo {volume} {35}},\ \bibinfo
  {pages} {557} (\bibinfo {year} {1926})}\BibitemShut {NoStop}%
\bibitem [{\citenamefont {Maggiore}(2011)}]{Maggiore:2010wr}%
  \BibitemOpen
  \bibfield  {author} {\bibinfo {author} {\bibfnamefont {M.}~\bibnamefont
  {Maggiore}},\ }\href {\doibase 10.1103/PhysRevD.83.063514} {\bibfield
  {journal} {\bibinfo  {journal} {Phys. Rev.}\ }\textbf {\bibinfo {volume}
  {D83}},\ \bibinfo {pages} {063514} (\bibinfo {year} {2011})},\ \Eprint
  {http://arxiv.org/abs/1004.1782} {arXiv:1004.1782 [astro-ph.CO]} \BibitemShut
  {NoStop}%
\bibitem [{\citenamefont {Casimir}(1948)}]{Casimir:1948dh}%
  \BibitemOpen
  \bibfield  {author} {\bibinfo {author} {\bibfnamefont {H.~B.~G.}\
  \bibnamefont {Casimir}},\ }\href@noop {} {\bibfield  {journal} {\bibinfo
  {journal} {Indag. Math.}\ }\textbf {\bibinfo {volume} {10}},\ \bibinfo
  {pages} {261} (\bibinfo {year} {1948})},\ \bibinfo {note} {[Kon. Ned. Akad.
  Wetensch. Proc.100N3-4,61(1997)]}\BibitemShut {NoStop}%
\bibitem [{\citenamefont {Lamoreaux}(1997)}]{Lamoreaux:1996wh}%
  \BibitemOpen
  \bibfield  {author} {\bibinfo {author} {\bibfnamefont {S.~K.}\ \bibnamefont
  {Lamoreaux}},\ }\href {\doibase 10.1103/PhysRevLett.81.5475,
  10.1103/PhysRevLett.78.5} {\bibfield  {journal} {\bibinfo  {journal} {Phys.
  Rev. Lett.}\ }\textbf {\bibinfo {volume} {78}},\ \bibinfo {pages} {5}
  (\bibinfo {year} {1997})},\ \bibinfo {note} {[Erratum: Phys. Rev.
  Lett.81,5475(1998)]}\BibitemShut {NoStop}%
\bibitem [{\citenamefont {Bordag}\ \emph {et~al.}(2001)\citenamefont {Bordag},
  \citenamefont {Mohideen},\ and\ \citenamefont
  {Mostepanenko}}]{Bordag:2001qi}%
  \BibitemOpen
  \bibfield  {author} {\bibinfo {author} {\bibfnamefont {M.}~\bibnamefont
  {Bordag}}, \bibinfo {author} {\bibfnamefont {U.}~\bibnamefont {Mohideen}}, \
  and\ \bibinfo {author} {\bibfnamefont {V.~M.}\ \bibnamefont {Mostepanenko}},\
  }\href {\doibase 10.1016/S0370-1573(01)00015-1} {\bibfield  {journal}
  {\bibinfo  {journal} {Phys. Rept.}\ }\textbf {\bibinfo {volume} {353}},\
  \bibinfo {pages} {1} (\bibinfo {year} {2001})},\ \Eprint
  {http://arxiv.org/abs/quant-ph/0106045} {arXiv:quant-ph/0106045 [quant-ph]}
  \BibitemShut {NoStop}%
\bibitem [{\citenamefont {Mohideen}\ and\ \citenamefont
  {Roy}(1998)}]{Mohideen:1998iz}%
  \BibitemOpen
  \bibfield  {author} {\bibinfo {author} {\bibfnamefont {U.}~\bibnamefont
  {Mohideen}}\ and\ \bibinfo {author} {\bibfnamefont {A.}~\bibnamefont {Roy}},\
  }\href {\doibase 10.1103/PhysRevLett.81.4549} {\bibfield  {journal} {\bibinfo
   {journal} {Phys. Rev. Lett.}\ }\textbf {\bibinfo {volume} {81}},\ \bibinfo
  {pages} {4549} (\bibinfo {year} {1998})},\ \Eprint
  {http://arxiv.org/abs/physics/9805038} {arXiv:physics/9805038 [physics]}
  \BibitemShut {NoStop}%
\bibitem [{\citenamefont {Roy}\ \emph {et~al.}(1999)\citenamefont {Roy},
  \citenamefont {Lin},\ and\ \citenamefont {Mohideen}}]{Roy:1999dx}%
  \BibitemOpen
  \bibfield  {author} {\bibinfo {author} {\bibfnamefont {A.}~\bibnamefont
  {Roy}}, \bibinfo {author} {\bibfnamefont {C.-Y.}\ \bibnamefont {Lin}}, \ and\
  \bibinfo {author} {\bibfnamefont {U.}~\bibnamefont {Mohideen}},\ }\href
  {\doibase 10.1103/PhysRevD.60.111101} {\bibfield  {journal} {\bibinfo
  {journal} {Phys. Rev.}\ }\textbf {\bibinfo {volume} {D60}},\ \bibinfo {pages}
  {111101} (\bibinfo {year} {1999})},\ \Eprint
  {http://arxiv.org/abs/quant-ph/9906062} {arXiv:quant-ph/9906062 [quant-ph]}
  \BibitemShut {NoStop}%
\bibitem [{\citenamefont {Bressi}\ \emph {et~al.}(2002)\citenamefont {Bressi},
  \citenamefont {Carugno}, \citenamefont {Onofrio},\ and\ \citenamefont
  {Ruoso}}]{Bressi:2002fr}%
  \BibitemOpen
  \bibfield  {author} {\bibinfo {author} {\bibfnamefont {G.}~\bibnamefont
  {Bressi}}, \bibinfo {author} {\bibfnamefont {G.}~\bibnamefont {Carugno}},
  \bibinfo {author} {\bibfnamefont {R.}~\bibnamefont {Onofrio}}, \ and\
  \bibinfo {author} {\bibfnamefont {G.}~\bibnamefont {Ruoso}},\ }\href
  {\doibase 10.1103/PhysRevLett.88.041804} {\bibfield  {journal} {\bibinfo
  {journal} {Phys. Rev. Lett.}\ }\textbf {\bibinfo {volume} {88}},\ \bibinfo
  {pages} {041804} (\bibinfo {year} {2002})},\ \Eprint
  {http://arxiv.org/abs/quant-ph/0203002} {arXiv:quant-ph/0203002 [quant-ph]}
  \BibitemShut {NoStop}%
\bibitem [{\citenamefont {Jaffe}(2005)}]{Jaffe:2005vp}%
  \BibitemOpen
  \bibfield  {author} {\bibinfo {author} {\bibfnamefont {R.~L.}\ \bibnamefont
  {Jaffe}},\ }\href {\doibase 10.1103/PhysRevD.72.021301} {\bibfield  {journal}
  {\bibinfo  {journal} {Phys. Rev.}\ }\textbf {\bibinfo {volume} {D72}},\
  \bibinfo {pages} {021301} (\bibinfo {year} {2005})},\ \Eprint
  {http://arxiv.org/abs/hep-th/0503158} {arXiv:hep-th/0503158 [hep-th]}
  \BibitemShut {NoStop}%
\bibitem [{\citenamefont {Lifshitz}(1956)}]{Lifshitz:1956zz}%
  \BibitemOpen
  \bibfield  {author} {\bibinfo {author} {\bibfnamefont {E.~M.}\ \bibnamefont
  {Lifshitz}},\ }\href@noop {} {\bibfield  {journal} {\bibinfo  {journal} {Sov.
  Phys. JETP}\ }\textbf {\bibinfo {volume} {2}},\ \bibinfo {pages} {73}
  (\bibinfo {year} {1956})}\BibitemShut {NoStop}%
\bibitem [{\citenamefont {Dzyaloshinskii}\ \emph {et~al.}(1961)\citenamefont
  {Dzyaloshinskii}, \citenamefont {Lifshitz},\ and\ \citenamefont
  {Pitaevskii}}]{0038-5670-4-2-R01}%
  \BibitemOpen
  \bibfield  {author} {\bibinfo {author} {\bibfnamefont {I.~E.}\ \bibnamefont
  {Dzyaloshinskii}}, \bibinfo {author} {\bibfnamefont {E.~M.}\ \bibnamefont
  {Lifshitz}}, \ and\ \bibinfo {author} {\bibfnamefont {L.~P.}\ \bibnamefont
  {Pitaevskii}},\ }\href {http://stacks.iop.org/0038-5670/4/i=2/a=R01}
  {\bibfield  {journal} {\bibinfo  {journal} {Soviet Physics Uspekhi}\ }\textbf
  {\bibinfo {volume} {4}},\ \bibinfo {pages} {153} (\bibinfo {year}
  {1961})}\BibitemShut {NoStop}%
\bibitem [{\citenamefont {Schwinger}\ \emph {et~al.}(1979)\citenamefont
  {Schwinger}, \citenamefont {DeRaad},\ and\ \citenamefont
  {Milton}}]{Schwinger:1977pa}%
  \BibitemOpen
  \bibfield  {author} {\bibinfo {author} {\bibfnamefont {J.~S.}\ \bibnamefont
  {Schwinger}}, \bibinfo {author} {\bibfnamefont {L.~L.}\ \bibnamefont
  {DeRaad}, \bibfnamefont {Jr.}}, \ and\ \bibinfo {author} {\bibfnamefont
  {K.~A.}\ \bibnamefont {Milton}},\ }\href {\doibase
  10.1016/0003-4916(78)90172-0} {\bibfield  {journal} {\bibinfo  {journal}
  {Annals Phys.}\ }\textbf {\bibinfo {volume} {115}},\ \bibinfo {pages} {1}
  (\bibinfo {year} {1979})},\ \bibinfo {note} {[,676(1977)]}\BibitemShut
  {NoStop}%
\bibitem [{\citenamefont {Gupta}(2002)}]{Gupta:2002qu}%
  \BibitemOpen
  \bibfield  {author} {\bibinfo {author} {\bibfnamefont {A.}~\bibnamefont
  {Gupta}},\ }\href@noop {} {\  (\bibinfo {year} {2002})},\ \Eprint
  {http://arxiv.org/abs/hep-th/0210069} {arXiv:hep-th/0210069 [hep-th]}
  \BibitemShut {NoStop}%
\bibitem [{\citenamefont {Ghilencea}\ and\ \citenamefont
  {Lee}(2005)}]{Ghilencea:2005hm}%
  \BibitemOpen
  \bibfield  {author} {\bibinfo {author} {\bibfnamefont {D.~M.}\ \bibnamefont
  {Ghilencea}}\ and\ \bibinfo {author} {\bibfnamefont {H.~M.}\ \bibnamefont
  {Lee}},\ }\href {\doibase 10.1088/1126-6708/2005/09/024} {\bibfield
  {journal} {\bibinfo  {journal} {JHEP}\ }\textbf {\bibinfo {volume} {09}},\
  \bibinfo {pages} {024} (\bibinfo {year} {2005})},\ \Eprint
  {http://arxiv.org/abs/hep-ph/0505187} {arXiv:hep-ph/0505187 [hep-ph]}
  \BibitemShut {NoStop}%
\bibitem [{\citenamefont {Perlmutter}\ \emph {et~al.}(1998)\citenamefont
  {Perlmutter} \emph {et~al.}}]{Perlmutter:1997zf}%
  \BibitemOpen
  \bibfield  {author} {\bibinfo {author} {\bibfnamefont {S.}~\bibnamefont
  {Perlmutter}} \emph {et~al.} (\bibinfo {collaboration} {Supernova Cosmology
  Project}),\ }\href {\doibase 10.1038/34124} {\bibfield  {journal} {\bibinfo
  {journal} {Nature}\ }\textbf {\bibinfo {volume} {391}},\ \bibinfo {pages}
  {51} (\bibinfo {year} {1998})},\ \Eprint
  {http://arxiv.org/abs/astro-ph/9712212} {arXiv:astro-ph/9712212 [astro-ph]}
  \BibitemShut {NoStop}%
\bibitem [{\citenamefont {Riess}\ \emph {et~al.}(1998)\citenamefont {Riess}
  \emph {et~al.}}]{Riess:1998cb}%
  \BibitemOpen
  \bibfield  {author} {\bibinfo {author} {\bibfnamefont {A.~G.}\ \bibnamefont
  {Riess}} \emph {et~al.} (\bibinfo {collaboration} {Supernova Search Team}),\
  }\href {\doibase 10.1086/300499} {\bibfield  {journal} {\bibinfo  {journal}
  {Astron. J.}\ }\textbf {\bibinfo {volume} {116}},\ \bibinfo {pages} {1009}
  (\bibinfo {year} {1998})},\ \Eprint {http://arxiv.org/abs/astro-ph/9805201}
  {arXiv:astro-ph/9805201 [astro-ph]} \BibitemShut {NoStop}%
\bibitem [{\citenamefont {Bahcall}\ \emph {et~al.}(1999)\citenamefont
  {Bahcall}, \citenamefont {Ostriker}, \citenamefont {Perlmutter},\ and\
  \citenamefont {Steinhardt}}]{Bahcall:1999xn}%
  \BibitemOpen
  \bibfield  {author} {\bibinfo {author} {\bibfnamefont {N.~A.}\ \bibnamefont
  {Bahcall}}, \bibinfo {author} {\bibfnamefont {J.~P.}\ \bibnamefont
  {Ostriker}}, \bibinfo {author} {\bibfnamefont {S.}~\bibnamefont
  {Perlmutter}}, \ and\ \bibinfo {author} {\bibfnamefont {P.~J.}\ \bibnamefont
  {Steinhardt}},\ }\href {\doibase 10.1126/science.284.5419.1481} {\bibfield
  {journal} {\bibinfo  {journal} {Science}\ }\textbf {\bibinfo {volume}
  {284}},\ \bibinfo {pages} {1481} (\bibinfo {year} {1999})},\ \Eprint
  {http://arxiv.org/abs/astro-ph/9906463} {arXiv:astro-ph/9906463 [astro-ph]}
  \BibitemShut {NoStop}%
\bibitem [{\citenamefont {Abe}\ \emph {et~al.}(1999)\citenamefont {Abe},
  \citenamefont {Hashida}, \citenamefont {Muta},\ and\ \citenamefont
  {Purwanto}}]{Abe:1999ts}%
  \BibitemOpen
  \bibfield  {author} {\bibinfo {author} {\bibfnamefont {H.}~\bibnamefont
  {Abe}}, \bibinfo {author} {\bibfnamefont {J.}~\bibnamefont {Hashida}},
  \bibinfo {author} {\bibfnamefont {T.}~\bibnamefont {Muta}}, \ and\ \bibinfo
  {author} {\bibfnamefont {A.}~\bibnamefont {Purwanto}},\ }\href {\doibase
  10.1142/S0217732399001097} {\bibfield  {journal} {\bibinfo  {journal} {Mod.
  Phys. Lett.}\ }\textbf {\bibinfo {volume} {A14}},\ \bibinfo {pages} {1033}
  (\bibinfo {year} {1999})},\ \Eprint {http://arxiv.org/abs/hep-ph/9905286}
  {arXiv:hep-ph/9905286 [hep-ph]} \BibitemShut {NoStop}%
\bibitem [{\citenamefont {Ghilencea}(2005)}]{Ghilencea:2004sq}%
  \BibitemOpen
  \bibfield  {author} {\bibinfo {author} {\bibfnamefont {D.~M.}\ \bibnamefont
  {Ghilencea}},\ }\href {\doibase 10.1088/1126-6708/2005/03/009} {\bibfield
  {journal} {\bibinfo  {journal} {JHEP}\ }\textbf {\bibinfo {volume} {03}},\
  \bibinfo {pages} {009} (\bibinfo {year} {2005})},\ \Eprint
  {http://arxiv.org/abs/hep-ph/0409214} {arXiv:hep-ph/0409214 [hep-ph]}
  \BibitemShut {NoStop}%
\bibitem [{\citenamefont {Ghilencea}\ \emph {et~al.}(2005)\citenamefont
  {Ghilencea}, \citenamefont {Hoover}, \citenamefont {Burgess},\ and\
  \citenamefont {Quevedo}}]{Ghilencea:2005vm}%
  \BibitemOpen
  \bibfield  {author} {\bibinfo {author} {\bibfnamefont {D.~M.}\ \bibnamefont
  {Ghilencea}}, \bibinfo {author} {\bibfnamefont {D.}~\bibnamefont {Hoover}},
  \bibinfo {author} {\bibfnamefont {C.~P.}\ \bibnamefont {Burgess}}, \ and\
  \bibinfo {author} {\bibfnamefont {F.}~\bibnamefont {Quevedo}},\ }\href
  {\doibase 10.1088/1126-6708/2005/09/050} {\bibfield  {journal} {\bibinfo
  {journal} {JHEP}\ }\textbf {\bibinfo {volume} {09}},\ \bibinfo {pages} {050}
  (\bibinfo {year} {2005})},\ \Eprint {http://arxiv.org/abs/hep-th/0506164}
  {arXiv:hep-th/0506164 [hep-th]} \BibitemShut {NoStop}%
\end{thebibliography}%
\bibliographystyle{apsrev4-1}

\end{document}